\documentclass[journal]{IEEEtran}
\ifCLASSINFOpdf

\else

\fi
\usepackage{graphicx}
\usepackage{subfig}
\usepackage{algorithm}
\usepackage[noend]{algorithmic}
\usepackage{amsmath}

\usepackage{bm}
\usepackage{amssymb}
\usepackage{amsmath}

\usepackage{amsfonts}
\usepackage{textcomp}
\usepackage{xcolor}
\usepackage{cite}
\usepackage{url}
\usepackage{booktabs}

\begin{document}

\title{Generative Memorize-Then-Recall framework for low bit-rate Surveillance Video Compression}

\author{Yaojun~Wu, Tianyu~He,
	Zhibo~Chen,~\IEEEmembership{Senior Member,~IEEE,}
        %John~Doe,~\IEEEmembership{Fellow,~OSA,}
        %and~Jane~Doe,~\IEEEmembership{Life~Fellow,~IEEE}% <-this % stops a space
\thanks{Zhibo Chen is the corresponding author. 

Yaojun~Wu and Zhibo~Chen are with the CAS Key Laboratory of Technology in Geo-spatial Information Processing and Application System, University of Science and Technology of China, Hefei 230027, China (e-mail: yaojunwu@mail.ustc.edu.cn, chenzhibo@ustc.edu.cn).
Tianyu~He is with the DAMO Academy in Alibaba Group (e-mail: timhe.hty@alibaba-inc.com).
}% <-this % stops a space
}


\markboth{Journal of \LaTeX\ Class Files,~Vol.~14, No.~8, August~2015}%
{Shell \MakeLowercase{\textit{et al.}}: Bare Demo of IEEEtran.cls for IEEE Journals}



\maketitle

\begin{abstract}
%Surveillance video applications grow dramatically in public safety and daily life, which often detect and recognize moving objects inside video signals. Existing surveillance video compression schemes are still based on traditional hybrid coding frameworks handling temporal redundancy by block-wise motion compensation mechanism, lacking the extraction and utilization of inherent structure information.In this paper, we alleviate this issue by decomposing surveillance video signals into the structure of a global spatio-temporal feature (memory) and skeleton for each frame (clue). The memory is abstracted by a recurrent neural network across Group of Pictures (GoP) inside one video sequence, representing appearance for elements that appeared inside GoP. While the skeleton is obtained by the specific pose estimator, it served as a clue for recalling memory. In addition, we introduce an attention mechanism to learn the relationship between appearance and skeletons. And we reconstruct each frame with an adversarial training process. Experimental results demonstrate that our approach can effectively generate realistic frames from appearance and skeleton accordingly. Compared with the latest video compression standard H.265, it shows much higher compression performance on surveillance video.

Applications of surveillance video have developed rapidly in recent years to protect public safety and daily life, which often detect and recognize objects in video sequences. Traditional coding frameworks remove temporal redundancy in surveillance video by block-wise motion compensation, lacking the extraction and utilization of inherent structure information. In this paper,  we figure out this issue by disentangling surveillance video into the structure of a global spatio-temporal feature (memory) for Group of Picture (GoP) and skeleton for each frame (clue).  The memory is obtained by sequentially feeding frame inside GoP into a recurrent neural network, describing appearance for objects that appeared inside GoP.  While the skeleton is calculated by a pose estimator, it is regarded as a clue to recall memory. Furthermore, an attention mechanism is introduced to obtain the relation between appearance and skeletons. Finally, we employ generative adversarial network to reconstruct each frame. Experimental results indicate that our method effectively generates realistic reconstruction based on appearance and skeleton, which show much higher compression performance on surveillance video compared with the latest video compression standard H.265.
\end{abstract}


% \fi
\IEEEpeerreviewmaketitle

\section{Introduction}

\IEEEPARstart{S}{urveillance} systems are widely applied for public safety, daily life, remote management, etc. Video sequences recorded by surveillance systems typically contain moving objects, especially human beings, and usually be used for detecting and recognizing. Considering the massive data generated by surveillance systems should be transmitted or stored, as well as being dealt with intelligent algorithms, urgent demands are imposed on an efficient and intelligent compression scheme.

\begin{figure}[t]
  \centering
  \includegraphics[width=0.45\textwidth]{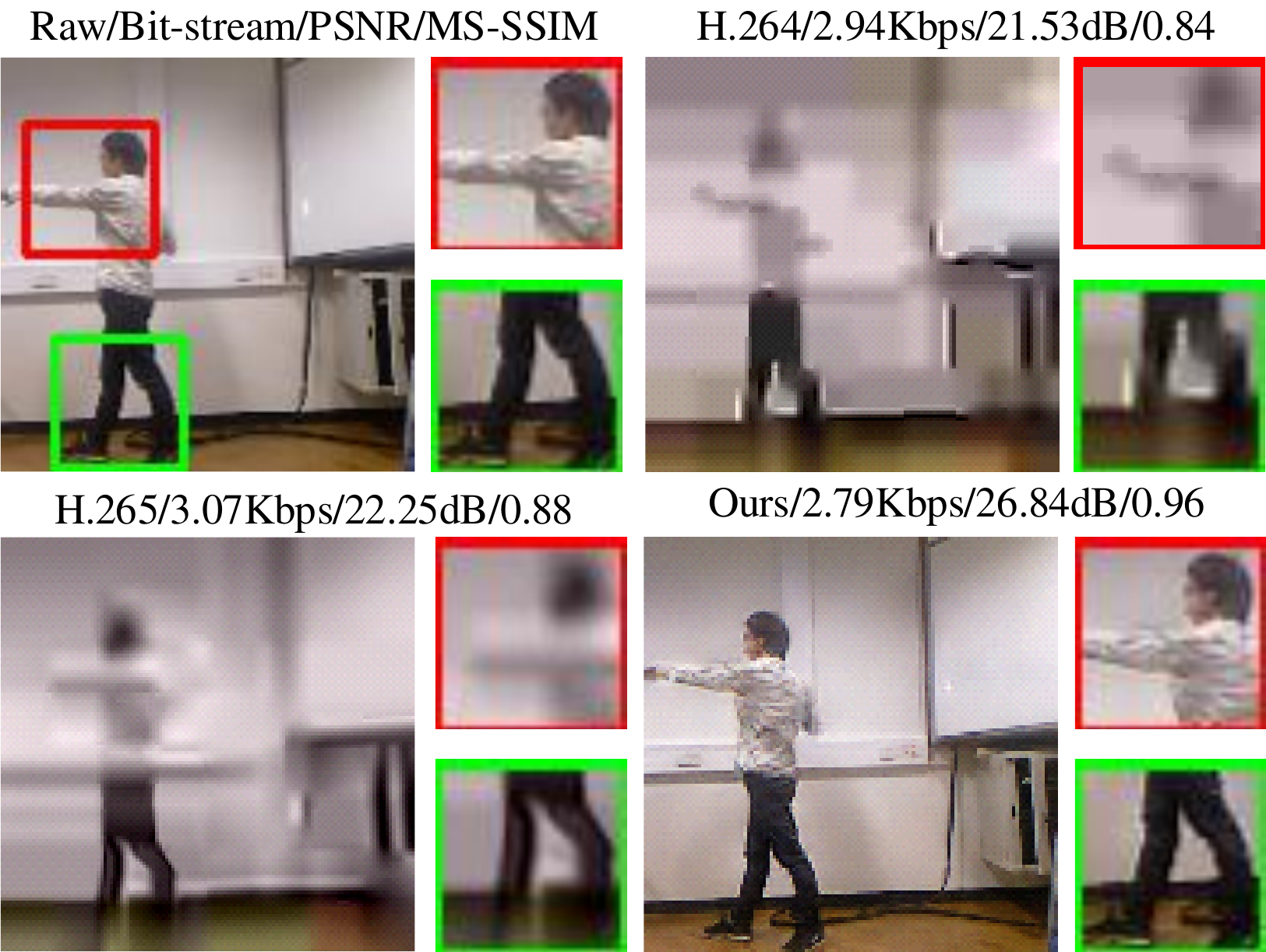}
  \caption{Visualization of reconstructed video frame compared to the latest video compression standards. It can be observed that our scheme achieves better reconstruction quality while using lower bit-rate.}
  \label{fig:firstpage}
\end{figure}

In general, the goal of the compression algorithm is to achieve a compact representation (bit-stream), from which the original content can be reconstructed in a lossy or lossless manner. Such an autoencoder-like process can be formulated as a Rate-Distortion Optimization (RDO) problem, where bit-rate and distortion (the original content v.s. the decompressed content) are considered in the optimization. Traditional hybrid video coding frameworks~\cite{habibi1974hybrid,forchheimer1981differential} (such as MPEG-2~\cite{itu1995generic}, H.264~\cite{wiegand2003overview} and H.265~\cite{sullivan2012overview}) typically tackle the compression problem with four basic steps: prediction, transform, quantization and entropy coding. In this kind of coding framework, the redundancy between neighboring frames is mainly de-correlated by block-wise motion compensation. Motion vectors are estimated by searching the best matching block from previous/subsequent frames, which is typically optimized for pixel-level fidelity (\textit{e.g.}, Mean Squared Error).

In the past two years, learning-based image compression has attracted wide attention~\cite{toderici2017full,rippel2017real,baig2017learning,balle2017end,theis2017lossy,mentzer2018conditional,minnen2018joint,he2018end}, while a few research works pay attention to end-to-end learned video compression. Wu \textit{et al.}~\cite{wu2018video} formulate the video compression as an interpolation problem, and reduce the redundancy in adjacent frames with pre-defined keyframes. Chen \textit{et al.}~\cite{chen2019learningvideo} propose a block-based compression scheme by modeling the spatio-temporal coherence. Lu \textit{et al.}~\cite{lu2018dvc} mimic traditional coding frameworks with learning-based components.

However, both traditional coding standards and learning-based schemes are typically optimized for pixel-level fidelity. The inherent structure information inside video signals is not well exploited during the compression process. In this paper, we want to formulate the video compression from the perspective of the semantic-level video decomposition. Our contributions can be summarized as follows.

\begin{itemize}
    \item We firstly decompose video compression problem as memorizing and recalling processes. Then we propose a new end-to-end video compression framework, named Memorize-Then-Recall (MTR).
     \item We provide a paradigm for semantic deep video compression. It leverages the success of variational autoencoder (VAE) and generative adversarial network (GAN). To best of our knowledge, this is the first VAE-GAN based end-to-end video compression framework.
%    \item With the learned inherent structure, we can also synthesis video where the appearance is abstracted from one video sequence and the motion is obtained from another unseen video sequence.
    \item We verify our MTR framework on video sequences with someone moving around or performing various actions, and achieve superior performance compared with the latest coding standards.
\end{itemize}

The remainder of this paper is organized as follows. Section~\ref{sec:hybrid} and Section~\ref{sec:learning} introduce the related work from the perspective of traditional hybrid coding framework and deep learning based compression, respectively. In Section~\ref{sec:II}, we formulate our VAE-GAN framework, and detailed description of our model is illustrated in Section~\ref{sec:mtr}. We will present experimental results in Section~\ref{sec:exps}, and then conclude in Section~\ref{sec:con}.
 
\section{Brief Introduction to Traditional Hybrid Coding Framework}
\label{sec:hybrid}
Hybrid Video Coding (HVC) has been widely adopted into most popular existing image/video coding standard like MPEG-2~\cite{itu1995generic}, H.264~\cite{wiegand2003overview} and H.265~\cite{sullivan2012overview}. Basically, it consists of prediction, transformation, quantization and entropy coding modules. We provided a short explanation of them in the following paragraphs.

\paragraph{\textbf{Prediction}}
Video prediction is introduced to remove the temporal redundancy in traditional video compression. Instead of coding an original pixel value directly, the residual between the original pixel and predicted pixel is encoded after prediction. Prediction typically consists of motion estimation and motion compensation~\cite{chen2006fast}. Motion estimation tries to search through all the possible blocks according to a specific matching criterion (\textit{e.g.}, Sum of Absolute Difference (SAD) or Mean Square Error (MSE)) to find the best matching blocks. While motion compensation directly copies pixels from the matched block in the previously reconstructed frame, and generate a predicted frame.

\paragraph{\textbf{Transform}}

Transform de-correlates coefficients to make them amendable to efficient entropy coding with low-order models. Besides, transform can distribute the signal energy to a few coefficients, which makes it easy to reduce redundancy and correlation. By performing on a reversible linear transform, signal can be transformed into frequency domain such as discrete cosine transform (DCT) and wavelet transform~\cite{xiong1999comparative}. 

\paragraph{\textbf{Quantization}}
Quantization is a many-to-one mapping that reduces the number of possible signal values while introducing some numerical errors in the reconstructed signal.
Quantization can be performed either on individual values (called scalar quantization) or on a group of values (called vector quantization)~\cite{nasrabadi1988image}. Typically, quantization can significantly reduce the amount of information that required to be transmitted.

\paragraph{\textbf{Entropy Coding}}
In entropy coding, variable-length code and arithmetic coding~\cite{rissanen1981universal} are common methods. Given source with non-uniform distribution, it can be compressed using a variable-length code where shorter code words are assigned to frequently occurring symbols and longer codes are assigned to less frequently occurring symbols~\cite{marpe2006h}. Besides, arithmetic coding compresses source by continuously dividing the probability interval of the input symbol.

\begin{figure*}[t]
\setlength{\abovecaptionskip}{0pt} 
\setlength{\belowcaptionskip}{-0pt}
	\centering
	\includegraphics[width=0.85\linewidth]{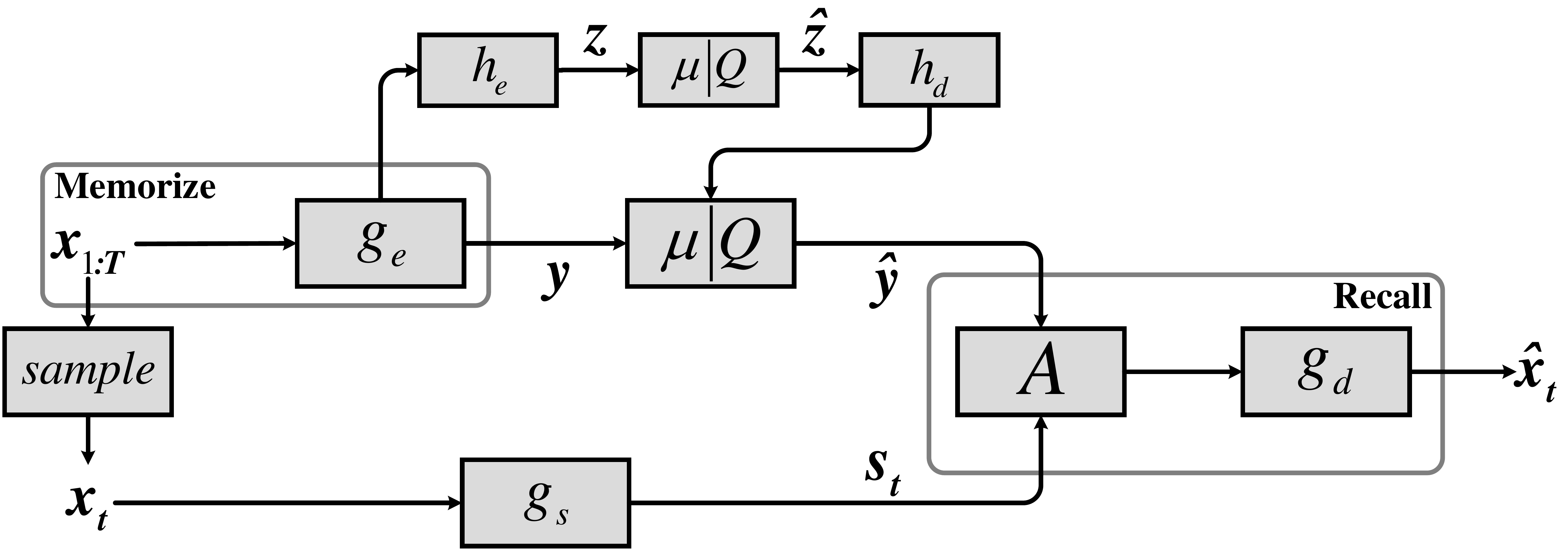}
	\caption{Diagram showing the operational structure of the video compression model. Boxes represent the operation, while arrows indicate the flow of data. Transformation operations ($g_e$, $g_d$, $h_e$, $h_d$, $g_s$) and attention operation ($A$) are described in section \ref{sec:II}. The operation labeled with $\mu \mid Q$ represents adding uniform noise during training phase, while it represents quantization and arithmetic coding in testing phase.}
	\label{fig:VAEGAN}
\end{figure*}

\section{Learning based Compression}
\label{sec:learning}
\subsection{Deep Image Compression}
In general, deep image compression method jointly trains all modules (transformation, quantization, entropy coding, etc.) with a goal to minimize the length of bit-stream as well as the distortion (e.g., PSNR or MS-SSIM~\cite{wang2004image}) of the reconstructed image. Among deep compression methods, Toderici \textit{et al.}~\cite{toderici2015variable,toderici2017full} utilize recurrent neural network (RNN) to compress the image by iteratively feeding the residual information (between reconstructed image and original image) into the RNN-based encoder. In their framework, the compression model can realize variable rates with a single network, without any retraining process.  To further improve the compression performance, a spatial adaptive bit allocation method \cite{johnston2018improved} is adopted to apply various bits on different locations for better reconstructed quality.

Besides RNN-based compression methods, variational autoencoder (VAE) also demonstrates its effectiveness in compression. The structure of the VAE-based compression method is firstly proposed in \cite{balle2016end}, which formulates their framework with non-linear transformation module (convolution with generalized divisive normalization \cite{balle2015density}) and relaxed quantization (uniform noise\cite{balle2016end}). Then their framework is evolved step by step with soft-to-hard vector quantization\cite{agustsson2017soft}, conditional entropy probability modeling incorporated with additional hyperprior information \cite{balle2018variational}, and Pixel-CNN based autoregressive context model in entropy modeling \cite{lee2018context,minnen2018joint}. To realize rate control, content-weighted importance-map is utilized in \cite{li2018learning}.

In addition to utilizing common distortion metric (PSNR or MS-SSIM), adversarial loss\cite{santurkar2018generative, agustsson2019generative} and semantic distortion metric (e.g., face verification accuracy distortion\cite{he2018end}) are investigated to improve the reconstructed quality from the perspective of human visual quality or certain task (e.g., face recognition) application.

\subsection{Deep Video Compression}
For video compression task, lots of works leveraging the success of artificial neural networks (ANN) to improve compression performance. They combine ANN with the traditional compression framework to improve the performance of one particular module, such as post-processing~\cite{lu2018deep}, mode decision~\cite{liu2016cu},  residual coding~\cite{chen2017deepcoder} and entropy coding~\cite{song2017neural}. 

In~\cite{chen2019learningvideo}, Chen \textit{et al.} propose a learning based video compression method that takes block as the basic processing unit. Wu \textit{et al.}~\cite{wu2018video} formulate the interpolation based compression method, which removes the redundancy between frames by using interpolation technique to predict frames. Han \textit{et al.}~\cite{han2018deep} utilize VAE model and introduce two branch encoder to obtain global and local information in the video, which is optimized through pixel-level fidelity. More recently, Lu \textit{et al.}~\cite{lu2018dvc} replace traditional video compression modules with neural network components. They first remove the temporal redundancy by predicting frames through the optical flow. After that, they compress the residual information through VAE based network. Rippel et al.~\cite{rippel2018learned} consider the redundancy between the optical flow and the residual information. They utilize concatenation to combine these information and feed them into VAE based framework for compression.

\section{The combination of VAE and GAN in Deep Video Compression}
\label{sec:II}
Different from traditional hybrid coding frameworks that heuristically optimize each component, we train our framework with end-to-end manner. We leverage the success of variational autoencoder (VAE)~\cite{pu2016variational} in image compression, and combine it with generative adversarial network (GAN)~\cite{goodfellow2014generative}. 
%Inspired by transform coding method in compression~\cite{goyal2001theoretical}, for a Group of pictures (GoP) $\bm{x_{1:T}}$ in video sequence, we employ a parametric transform $g_{e}(\bm{x_{1:T}},\bm{\phi_{ga}})$ to transform it into latent representation $\bm{M}$. Since the $\bm{M}$ contains the global information of one GoP, which is similar to the mechanism of human memory, we also call it memory in our paper. The memory information $\bm{M}$ is then quantized to form $\bm{\hat{M}}$, which will be losslessly compressed by entropy coding techniques (arithemetic coding~\cite{rissanen1981universal}).  
Detailed information about the combination is described in the following subsections. 
  
\subsection{Variational Autoencoder (VAE) based Deep Video Compression}
\label{sec:vae}
We first solve the video compression problem by VAE. VAE has demonstrated its effectiveness in deep image compression method~\cite{balle2018variational, minnen2018joint}, which even achieves better performance with BPG. Here we expand it into video compression task. 

For a Group of pictures (GoP) $\bm{x_{1:T}}$ in video sequence, we employ a parametric transform $g_{e}(\bm{x_{1:T}},\bm{\phi_{g_e}})$ to transform it into latent representation $\bm{M}$. Since the $\bm{M}$ contains the global information of one GoP, which is similar to the mechanism of human memory, we also call it memory in our paper. The memory information $\bm{M}$ is then quantized to form $\bm{\hat{M}}$, which will be losslessly compressed by entropy coding techniques (arithemetic coding~\cite{rissanen1981universal}). 

In order to further remove the spatial redundancy in quantized memory $\bm{\hat{M}}$, we following Ball{\'e} \textit{et al.}~\cite{balle2018variational} to utilize hyperprior $z$ to predict the probability of $\bm{\hat{M}}$ in entropy coding, which is obtained from the hyperprior parametric transformation $h_{e}(\bm{M}; \bm{\phi_{h_{e}}})$. Then we feed $\bm{\hat{M}}$ into the parametric transformation $g_{d}(\bm{\hat{M}},\bm{\phi_{g_{d}}})$ to obtain the reconstructed GoP $\bm{\hat{x}_{1:T}}$. 

The goal of our VAE is to approximate the true posterior $p_{\bm{\hat{y}} \mid \bm{x_{1:T}}}(\bm{\hat{y}} \mid \bm{x_{1:T}})$ with a parametric variational density $q(\bm{\hat{M}},\bm{\hat{z}}\mid \bm{x_{1:T}})$ by minimizing their Kullback-Leibler (KL) distance over the data distribution $p_{\bm{x_{1:T}}}$:
\begin{equation}
\begin{aligned}
\mathbb{E}_{\bm{x_{1:T}}\sim p_{\bm{x_{1:T}}}}D_{KL}[q(\bm{\hat{M}},\bm{\hat{z}}\mid \bm{x_{1:T}})\parallel p_{\bm{\hat{M}},\bm{\hat{z}}\mid \bm{x_{1:T}}}(\bm{\hat{M}},\bm{\hat{z}}\mid \bm{x_{1:T}})]\\
=\mathbb{E}_{\bm{x_{1:T}}\sim p_{\bm{x_{1:T}}}}\mathbb{E}_{\bm{\hat{M}},\bm{\hat{z}}\sim q}[
\underbrace{\log q(\bm{\hat{M}},\bm{\hat{z}}\mid \bm{x_{1:T}})}_{0}
\underbrace{-\log p(\bm{x_{1:T}}\mid \bm{\hat{M}})}_{D(distortion)}\\
\underbrace{-\log p_{\bm{\hat{M}} \mid \bm{\hat{z}}}(\bm{\hat{M}} \mid \bm{\hat{z}})
-\log p_{\bm{\hat{z}}}(\bm{\hat{z}})}_{R(rate)}]
+const.
\label{formula:kl}
\end{aligned}
\end{equation}

We extend the KL distance through Bayes' theorems in (\ref{formula:kl}). The final result contains three parts. The first part corresponds to the joint distribution of the quantized memory $\hat{M}$ and quantized hyperprior $z$, which is obtained through parametric transformation and adding uniform noise ( as a  substitution of quantization~\cite{balle2017end}). Hence, the first part can be written as follows: 
\begin{equation}
\begin{aligned}
&q(\bm{\hat{M}},\bm{\hat{z}}\mid \bm{x_{1:T}},\bm{\phi _{g_{e}}}, \bm{\phi_{h_{e}}})=\\
&\prod_{i}\mu(\hat{M_{i}} \mid M_{i}-\frac{1}{2}, M_{i}+\frac{1}{2})\times
\prod_{j}\mu(\hat{z_{j}} \mid z_{j}-\frac{1}{2}, z_{j}+\frac{1}{2})\\
&\quad \quad \quad \quad with\ \bm{M}=g_{e}(\bm{x_{1:T}},\bm{\phi_{g_{e}}}), \bm{z}=h_{e}(\bm{M};\bm{\phi_{h_{e}}}),
\label{formula:M}
\end{aligned}
\end{equation}
where $\mu$ denotes the uniform distribution centered on $M_{i}$ or $z_{i}$, $\hat{z_{i}}$ denotes the quantized hyperprior and $\bm{\phi}$ denotes the corresponding parameters of network. Since the width of the uniform distribution is constant, the result of the first part is equal to zero. Therefore, the first part can be ignored in our loss function. 

The second part corresponds to the distortion. In VAE based compression method, this part is always given by Gaussian distribution $N(\bm{x_{1:T}} \mid \bm{\hat{x}_{1:T}}, (2\lambda)^{-1})$, which is equal to the squared difference between input GoP $\bm{x_{1:T}}$ and reconstructed GoP $\bm{\hat{x}_{1:T}}$. However, this kind of distortion metric can only measure the distortion from the perspective of pixel level fidelity, which is inconsistent with the human visual system. Hence, we propose to introduce generative adversarial network in VAE framework, which will be detailedly discussed in section \ref{sec:gan}.

The third part represents the total rate of the encoding in VAE-based compression. It includes two items, namely the transmission cost of quantized memory $\bm{\hat{M}}$ and quantized hyperprior $\bm{\hat{z}}$. For each element $\hat{M_{i}}$ in quantized memory $\bm{\hat{M}}$,  we assume it follows a zero-mean Gaussian distribution, in which the standard deviation is predicted by quantized hyperprior $\hat{z}$ and parametric transform $h_{d}(\bm{\hat{z}},\bm{\theta_{h_{d}}})$. Therefore, the rate of quantized memory $\bm{\hat{M}}$ can be written as follows:
\begin{equation}
\begin{aligned}
p_{\bm{\hat{M}} \mid \bm{\hat{z}} }(\bm{\hat{M}} \mid \bm{\hat{z}},\phi_{h_{e}})=\prod_{i}(N(0,\sigma_{i}^{2})*\mu(-\frac{1}{2},\frac{1}{2}))(\hat{M_{i}})\\
with\ \bm{\hat{\sigma}}=h_{d}(\bm{\hat{z}},\bm{\theta_{h_d}}).
\end{aligned}
\end{equation}

As the hyperprior have no prior to predict its density, here we follow Ball{\'e} \textit{et al.}~\cite{balle2018variational} utilizing a non-parametric, fully factorized model to predict its probability, which can be seen as follows:
\begin{equation} 
\begin{aligned}
p_{\bm{\hat{z}}}(\bm{\hat{z}})=\prod_{i}(p_{z_{i}}*\mu(-\frac{1}{2},\frac{1}{2}))(\hat{z_{i}}).
\end{aligned}
\end{equation}

In the above VAE-based framework, we utilize four parametric transforms ($g_{e}$, $g_{d}$, $h_{e}$ and $h_{d}$) to realize the compress and decompress procedures. In theoretical, these parametric transforms can be any parameterized function. In this paper, we utilize artificial neural networks as these transforms, and the detailed structure will be discussed in section \ref{sec:mtr}.

\subsection{Introduce GAN into VAE based Compression Framework}
\label{sec:gan}
Generative models like Generative Adversarial Networks (GANs) achieve impressive success in lots of tasks recently. In a typical scene, GANs consist of a generator and a discriminator. 
The core of the GAN is to optimize the minimax game between generator and discriminator. The discriminator aims to determine whether the input is real data, while the goal of the generator is to generate as much realistic data as possible to deceive the discriminator.
%The generator is responsible for generating realistic data that can not be distinguished from real data, while the discriminator simultaneously learns to distinguish the generated data from the real distribution. 
Such an adversarial training scheme facilitates the generator to yield the generated data with the same distribution as real data. In 2014, Mirza \textit{et al.}~ \cite{mirza2014conditional} further extend GANs into a conditional version, in which some extra information is used as condition when generating data. 

Similarly, we treat the process of reconstruction $g_{d}$ in VAE as a kind of conditional generation~\cite{mirza2014conditional}, which can be seen in Fig.\ref{fig:VAEGAN}. 
%Similarly, we treat the process of reconstruction as a kind of conditional generation~\cite{mirza2014conditional}, which is combined with our recalling attention mechanism. 
%where the generation is conditioned on the output of Recalling Attention. We base our network of the generator on the objective presented in pix2pixHD~\cite{wang2018vid2vid}. The goal of the generator is to yield reconstructed frames that cannot be distinguished by two discriminators:
Specifically, for $t\textit{-}th$ frame $\bm{x_{t}}$ in GoP $\bm{x_{1:T}}$, we first introduce per-frame information $\bm{s_t}$ to help the reconstruction, which is obtained by network $g_{s}(\bm{x_t}, \bm{\phi_{s}})$. The local information $\bm{s_t}$  is also called clue in this paper, it helps the reconstruction of each frame. Then we propose recalling attention mechanism $A(\bm{\hat{M}}, \bm{s_t}, \bm{\phi_{A}})$ to combine the global information $\bm{\hat{M}}$ and local information $\bm{s_t}$, which will be discussed in section \ref{sec:recall}. After that, the joint representation is fed into the generator $g_d$ to obtain the reconstructed frame $\bm{\hat{x}_{t}}$. 

\begin{algorithm}[t]  
  \caption{MTR surveillance video compression framework.}   
  \label{alg:MTR}  
  \begin{algorithmic}[1]
    \REQUIRE \quad \\
    		      The input GoP $\bm{x}_{1:T}$;\\
    		      Training: Flag (1 for training and 0 for testing);
    \ENSURE Reconstructed GoP $\bm{\hat{x}}_{1:T}$ and Bitstream $B$.
    \STATE $B \xleftarrow{} [\ ]$;
    \STATE $\bm{M} \xleftarrow{} g_{e}(\bm{x_{1:T}},\bm{\phi_{g_{e}}})$; 
    \STATE $z \xleftarrow{} h_e(\bm{M},\bm{\phi_{h_e}})$
    \IF{Training}
        \STATE $\bm{\hat{M}} \xleftarrow{} \bm{M}+\mu(-\frac{1}{2},\frac{1}{2})$; 
        \STATE $\bm{\hat{z}} \xleftarrow{} \bm{z}+\mu(-\frac{1}{2},\frac{1}{2})$; 
    \ELSE
        \STATE $\bm{\hat{M}} \xleftarrow{} round(\bm{M})$;
        \STATE $\bm{\hat{z}} \xleftarrow{} round(\bm{z})$;
    \ENDIF
    \STATE $p_{\bm{\hat{M}}} \xleftarrow{} h_{d}(\bm{\hat{z}},\phi_{h_d})$
    \STATE $B \xleftarrow{} Concat(B,arithmetic\ coding(\bm{\hat{M}},p_{\bm{\hat{M}}}))$
    \STATE $B \xleftarrow{} Concat(B,arithmetic\ coding(\bm{\hat{z}}))$
    \FOR{t \textless T}
    	\STATE $s_t \xleftarrow{} g_s(\bm{x_t},\bm{\phi_s})$
	\STATE $B \xleftarrow{} Concat(B,lossless\ coding(s_t))$
	\STATE Joint\ Feature $\bm{F}\xleftarrow{} A(\bm{\hat{M}}, \bm{s_t}, \bm{\phi_A})$
	\STATE $\bm{\hat{x}_t} \xleftarrow{} g_d(\bm{F}, \bm{\phi_{g_d}})$
    \ENDFOR
    \STATE $\bm{\hat{x}_{1:T}} \xleftarrow{} Concate(\bm{\hat{x}_1}, \ldots, \bm{\hat{x}_T})$
    \STATE Return $\bm{\hat{x}_{1:T}}$, $B$ 
  \end{algorithmic}  
\end{algorithm}

For discriminator, it needs to determine as accurately as possible whether the input is real or generated, which can be optimized through the following formula:
\begin{equation}
\begin{aligned}
L_{dis_S}=
\mathbb{E}_{\bm{x_{1:T} \sim p_{\bm{x_{1:T}}}}}[\log \prod_{t=1}^{T} p_{d_S}(1 \mid \bm{x_{t}})]+\\
\mathbb{E}_{\bm{\hat{x}_{1:T} \sim p_{\bm{\hat{x}_{1:T}}}}}[\log \prod_{t=1}^{T} p_{d_S}(0 \mid \bm{\hat{x}_{t}})]\\
with \ p_{d_S}(1\mid \bm{x_{t}})=g_{dis_S}(\bm{x_t}, \bm{s_{t}}, \bm{\phi_{dis_S}}),\\
p_{d_S}(0\mid \bm{\hat{x}_{t}})=1-g_{dis_S}(\bm{\hat{x_t}}, \bm{s_{t}} ,\bm{\phi_{dis_S}}),\\
\bm{s_t}=g_{s}(\bm{x_t}, \bm{\phi_{s}}),
\label{formula:dis}
\end{aligned}
\end{equation}
where the $p_{d_S}$ is the probability obtained from discriminator, predicting whether the input frame ($\bm{x_{t}}$ or $\bm{\hat{x}_{t}}$) is real data (1 for real and 0 for fake). In (\ref{formula:dis}), we assume that the generated frames are independent of each other, and decompose the judgment probability on GoP into the accumulation of judgment probability on each frame. Since the discriminator only judges whether the input is a real frame, we also call it spatial discriminator in this paper.  

For generator, it aims to generate frame as closer to real frame as possible. Thus, the loss of the generator can be written as:
%D的loss 引出G的loss，然后将loss扩展成两个D的情况
\begin{equation}
\begin{aligned}
L_{G}=-\mathbb{E}_{\bm{\hat{x}_{1:T} \sim p_{\bm{\hat{x}_{1:T}}}}}[\log \prod_{t=1}^{T} p_{d_{S}}(0 \mid \bm{\hat{x}_{t}})]\\
with \ \bm{\hat{x}_{t}}=g_{d}(A(\bm{\hat{M}}, \bm{s_t}, \bm{\phi_{A}}), \bm{\phi_{g_{d}}}).
\label{formula:G}
\end{aligned}
\end{equation}

As described in section \ref{sec:vae}, the distortion part in the VAE optimization function is always defined with square difference, which is inconsistent with the human visual system. In this part, we utilize the generator loss $L_G$ as the distortion function in (\ref{formula:kl}), which is more similar to the human visual system compared with the pixel-level fidelity. 
Based on the above combination of VAE and GAN, we can describe the detailed algorithm of our proposed MTR framework in Algorithm.~\ref{alg:MTR}.

In the above VAE-GAN based compression framework, we introduce local information $\bm{s_t}$ to improve the generation of each frame $\bm{x_t}$. Theoretically, local information $\bm{s_t}$ can be any feature. In this paper, we utilize the pose information (skeleton) obtained from pose estimator as the clue information.

\section{Detailed Framework of MTR}
\label{sec:mtr}

\begin{figure}[t]
\setlength{\abovecaptionskip}{0pt} 
\setlength{\belowcaptionskip}{-0pt}
	\centering
	\includegraphics[width=1.0\linewidth]{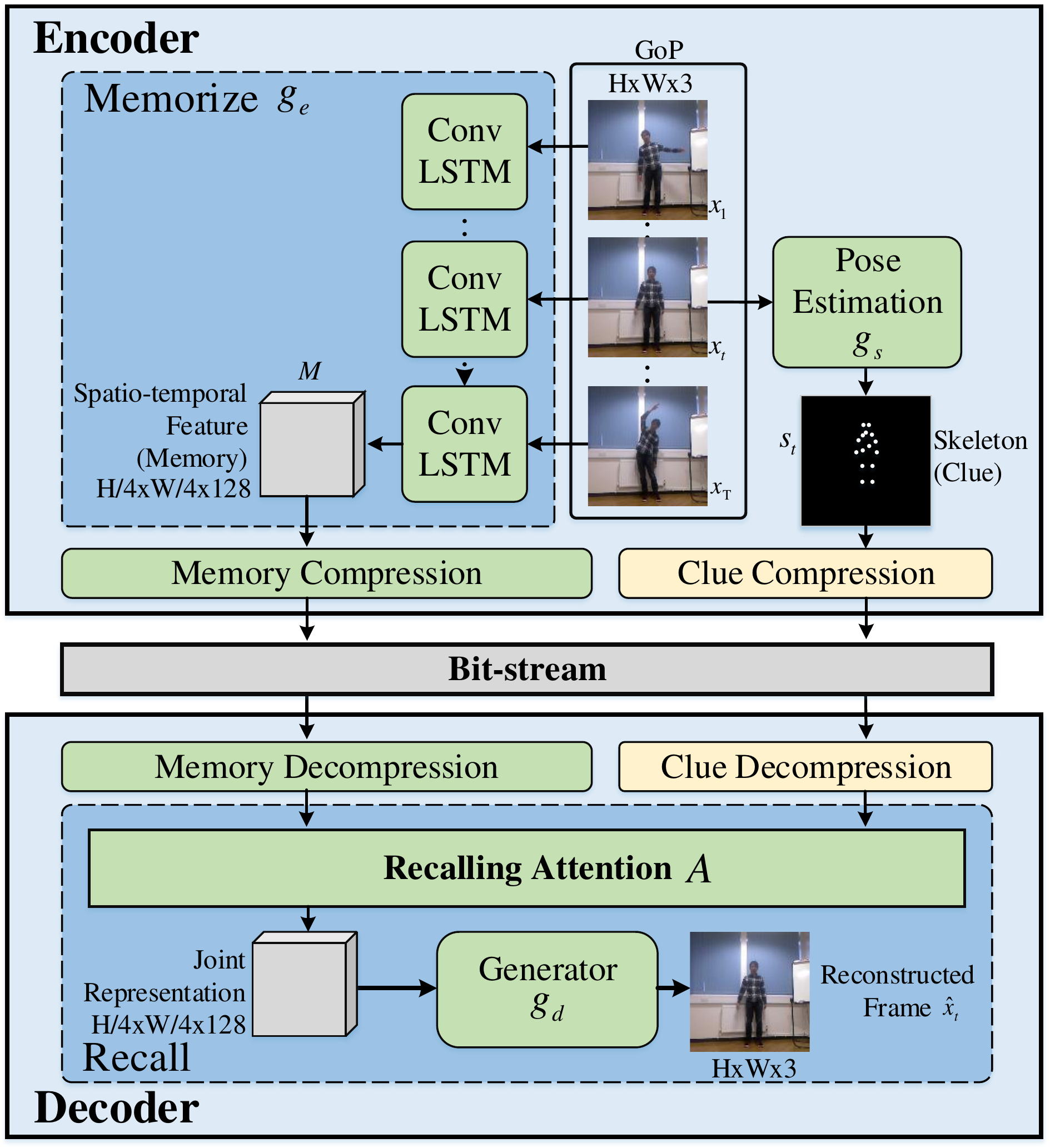}
	\caption{Memorize-Then-Recall (MTR) Framework. The top and bottom parts demonstrate the encoder and decoder respectively. The size of the feature is denoted as: height$\times$weight$\times$dimension. We jointly train Green modules with the loss defined in section~\ref{sec:loss}.}
	\label{fig:framework}
\end{figure}

In section~\ref{sec:II}, we formulate our video compression framework (MTR) through the combination of VAE and GAN. Here we give a detailed description of our MTR framework. The overall pipeline of the proposed MTR is illustrated in Fig.~\ref{fig:framework}.

%Different from traditional hybrid coding frameworks that heuristically optimize each component, we trained our framework with end-to-end manner. The overall pipeline of the proposed MTR is illustrated in Fig.~\ref{fig:framework}.

%For one video sequence, given a Group of Pictures (GoP) $\bm{x_{1:T}}}$, we decompose the video content into a global spatio-temporal feature $M$ (memory) and skeletons $S=\{s_1, s_2, \ldots, s_N\}$ (clues) for all frames that can be efficiently compressed, where $s_t$ denotes the skeleton extracted from $t\textit{-}th$ frame. 
For one video sequence, given a GoP $\bm{x_{1:T}}$, we decompose the video content into a global spatio-temporal feature $M$ (memory) and skeletons $\bm{s_{1:T}}$ (clues) for all frames that can be efficiently compressed. 

In the encoder, we utilize the Conv-LSTM as the encoder transformation $g_e$ and abstract the global spatio-temporal feature $M$ from GoP $\bm{x_{1:T}}$, standing for \textit{memory} to the GoP. It represents appearance for elements that appeared inside GoP, which will be further compressed by quantization and entropy coding ($h_e$ and $h_d$). For skeletons, they are obtained by the specific pose estimator~\cite{cao2017realtime}, which served as \textit{clues}. It will be compressed through predictive coding and entropy coding. 

%In the encoder, we abstract the global spatio-temporal feature $M$ by a recurrent neural network across GoP, standing for \textit{memory} to the GoP. It represents appearance for elements that appeared inside GoP, which will be further compressed by quantization and entropy coding. For skeletons, they are obtained by the specific pose estimator~\cite{cao2017realtime}, which served as \textit{clues}. It will be compressed through predictive coding and entropy coding. 

In the decoder, the reconstructed spatio-temporal feature $\bm{\hat{M}}$ and reconstructed skeletons $\bm{\hat{s}_{1:T}}$ can be obtained by corresponding inverse operations in the decompression phase. After that, we introduce a Recalling Attention mechanism $A$ to implement the recalling process, from which we can attain a feature that combines the information from the $\bm{\hat{M}}$ and $\bm{\hat{s_t}}$, describing the appearance with regard to the current frame. We then feed the joint representation into a generator and train it in conjunction with two different discriminators to achieve a realistic frame reconstruction $\bm{\hat{x_t}}$.

Detailed description of each component is in the following subsections.

\subsection{Memorize over Sequence}
\label{sec:memorize}
Typically, there exist high spatio-temporal correlations between pixels in a video sequence. Existing motion compensation (prediction) method is generally based on the assumption that each block in the current frame is related to a block of a previous/subsequent frame by the motion of objects. Therefore, they de-correlate highly correlated neighboring signal samples by directly copying the corresponding pixels according to estimated motion vectors. However, such pixel-level fidelity can not reflect the inherent structure information of objects. 

Hence, we leverage ConvLSTM~\cite{xingjian2015convolutional} to model spatio-temporal coherence inside GoP, which aims to obtain global information for one GoP. ConvLSTM utilizes a memory cell $\bm{C_t}$ as an accumulator of the state information. The cell is accessed, written and cleared by several self-parameterized controlling gates. Every time a new input comes, its information will be accumulated to the cell if the input gate $\bm{i_t}$ is activated. Also, the past cell status $\bm{c_{t-1}}$ could be "forgotten" in this process if the forget gate $\bm{f_t}$ is on. Whether the latest cell output $\bm{c_t}$ will be propagated to the final state $\bm{h_t}$ is further controlled by the output gate $\bm{o_t}$. In our method, We utilize frames ($\bm{x_{1:T}}$) in the GoP as the input of the ConvLSTM, and $C_t$ is the output of the ConvLSTM. The key equations of ConvLSTM are shown in the following:
\begin{equation}
% \begin{array}{1}
\begin{aligned}
    \bm{i_t}=& \sigma (\bm{W_{xi}}\ast \bm{x_t}+\bm{W_{hi}}\ast \bm{H_{t-1}} + \bm{W_{ci}}\circ \bm{C_{t-1}}+\bm{b_i}) \\
    \bm{f_t}=& \sigma (\bm{W_{xf}}\ast \bm{x_t}+\bm{W_{hf}}\ast \bm{H_{t-1}} + \bm{W_{cf}}\circ \bm{C_{t-1}}+\bm{b_f}) \\
    \bm{C_t}=& \bm{f_t}\circ \bm{C_{t-1}}+\bm{i_t}\circ tanh(\bm{W_{xc}}\ast \bm{x_t}+\bm{W_{hc}}\ast \bm{H_{t-1}} + \bm{b_{c}}) \\
    \bm{o_t}=& \sigma (\bm{W_{xo}}\ast \bm{x_t}+\bm{W_{ho}}\ast \bm{H_{t-1}} + \bm{W_{co}}\circ \bm{C_{t}}+\bm{b_o}) \\
    \bm{H_t}=& \bm{o_t}\circ tanh(\bm{C_t}),
\end{aligned}
% \end{array}
    \label{conv-lstm}
\end{equation}
where '$\ast$' denotes the convolution operator and '$\circ$' denotes the Hadamard product.

Specifically, we split a GoP into frames $\{\bm{x_1}, \bm{x_2}, \ldots, \bm{x_T}\}$. Then we sequentially feed them into ConvLSTM, and final output $\bm{C_T}$ is the global spatio-temporal feature for the whole GoP, which is leveraged as memory in our framework.

\begin{figure}[b]
    \setlength{\abovecaptionskip}{0pt} 
    \setlength{\belowcaptionskip}{-0pt}
    \centering
	\includegraphics[width=1.00\linewidth]{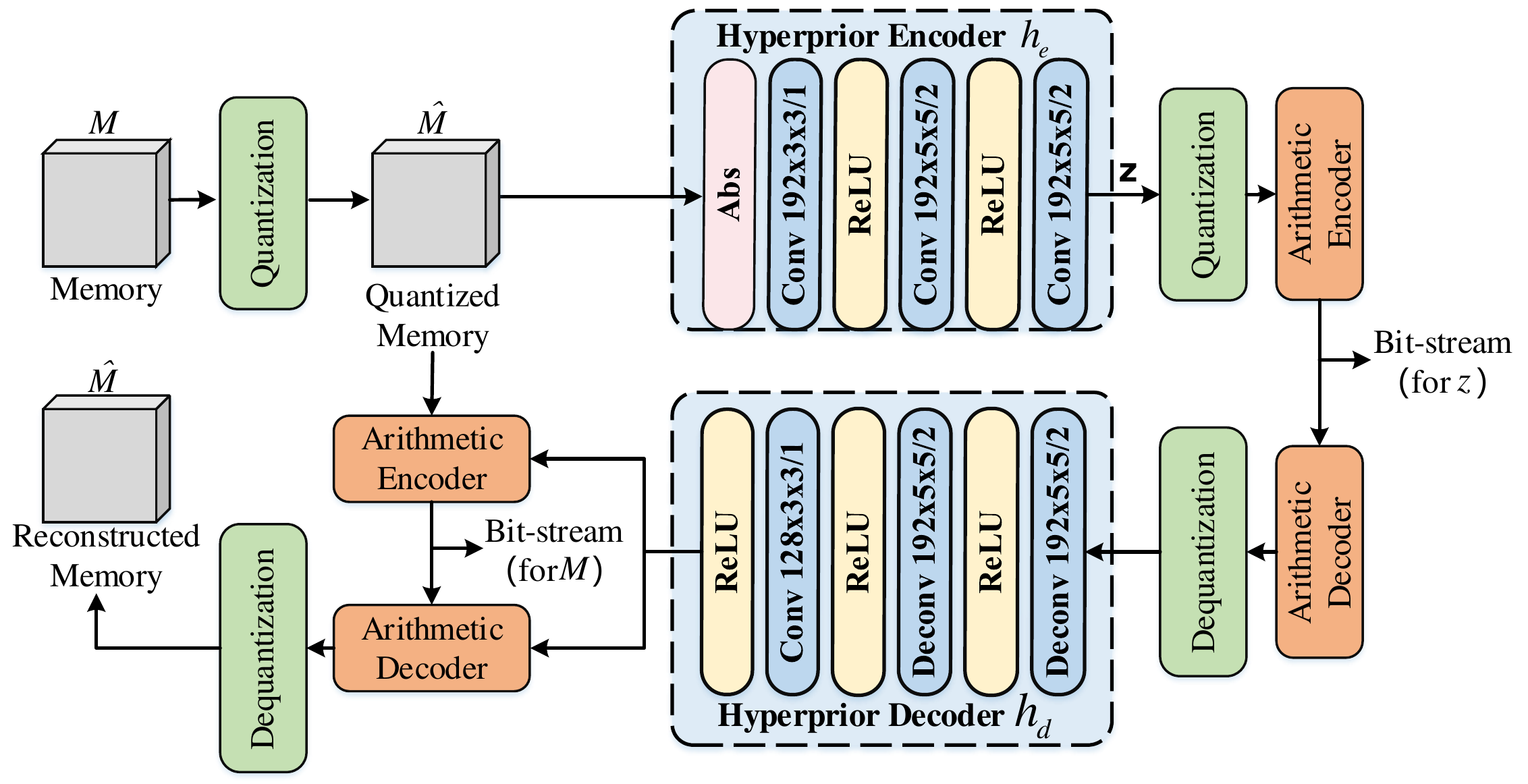}
	\caption{Our memory compression $\&$ decompression. Convolution parameters are denoted as: number of filters $\times$ kernel height $\times$ kernel width / stride.}
	\label{fig:memorycompress}
\end{figure}

\subsection{Memory Compression  $\&$ Decompression}
We utilize the spatio-temporal feature $\bm{M}$ as memory to represent appearance for elements that appeared inside GoP. To compress the spatio-temporal feature $\bm{M}$, we firstly apply a quantization operation. Then, the quantized spatio-temporal feature $\bm{\hat{M}}$ is fed into an entropy coder, which further reduces the redundancy with the help of the hyperprior network. Details about the quantization and entropy coding are stated as follows:

\paragraph{\textbf{Quantization}}
As described in (\ref{formula:M}), we utilize additive uniform noise during the training process. Formally, let $\mu(a,b)$ denote the uniform distribution on the interval $(a,b)$, the quantized spatio-temporal feature $\hat{M}$ in the training process can be approximated by:
\begin{equation}
\setlength{\abovedisplayskip}{3pt}
\setlength{\belowdisplayskip}{3pt}
    \hat{M}=M+\mu(-\frac{1}{2},\frac{1}{2}).
    \label{noise}
\end{equation}
%We utilize round operation as our quantization in the end-to-end MTR framework. However, the round operation is non-differentiable, which places an obstacle for end-to-end training. Inspired by~\cite{balle2016end}, we substitute round with an additive uniform noise during training. Formally, let $U(a,b)$ denote the uniform distribution on the interval $(a,b)$, the quantized spatio-temporal feature $\hat{M}$ in the training process can be approximated by:

Note that, such approximation is only performed in the training phase, while we directly apply rounding operation in the testing stage. By performing quantization, the feature memory $\bm{M}$ is successfully converted into a limited discrete representation, introducing a great reduction of bit-rate.

\paragraph{\textbf{Entropy Coding with Hyperprior Modeling}}

Context-based entropy coding is a general lossless compression method and commonly used after quantization in traditional coding frameworks. In theory, entropy coding can achieve an optimal solution with a known input probability distribution. However, for media content, the distribution of each sample is different from each other. This motivates some context-based coding scheme that automatically updates probability distribution according to encoded data. Similar to context-based entropy coding adapting to certain content, we introduce a hyperprior network ($h_e$, $h_d$) to predict the probability distribution for $\bm{\hat{M}}$ in (\ref{formula:kl}), which is illustrated in Fig.~\ref{fig:memorycompress}. 
From Fig.~\ref{fig:memorycompress}, we can see that the bitstream of quantized memory contains two parts, namely the rate for the hyperprior $\bm{\hat{z}}$ and the rate for the quantized memory $\bm{\hat{M}}$.
%\cite{balle2018variational} introduce a hyperprior $z$ to predict the probability distribution for current input content, while the predicted distribution is used for improving the efficiency of entropy coding. In this paper, we leverage this technique and construct a hyperprior network, which is illustrated in Fig.~\ref{fig:memorycompress}. 
%The hyperprior network takes the quantized spatio-temporal feature $\hat{M}$ as input to obtain hyperprior $z$, which will be utilized to predict a probability distribution $p_{\hat{M}}$. Based on $p_{\hat{M}}$, we can utilize entropy coding to compress $\hat{M}$. Since the $p_{\hat{M}}$ is needed in the entropy decoding process, hyperprior $z$ is encoded by quantization and entropy coding and transmitted along with the quantized memory.
%Following Ball\'{e} \textit{et al.}~\cite{balle2018variational}, we utilize the hyperprior network to predict the probability distribution $p_{\hat{M}}$ of $\hat{M}$,
%Therefore, the bit-stream from the memory compression includes two parts, $\hat{M}$ and $z$. We can estimate the bits by the following function:
%\begin{equation}
%\setlength{\abovedisplayskip}{3pt}
%\setlength{\belowdisplayskip}{3pt}
%    R_{\hat{M}} = \underbrace{\mathbb{E}_{\hat{M}\sim p_{\hat{M}}}[-\log_2 p_{\hat{M}}(\hat{M})]}_{rate(\hat{M})} + \underbrace{\mathbb{E}_{\hat{M}\sim p_{\hat{M}}}[-log_2 p_z(z)]}_{rate(z)}.\label{rate}
%\end{equation}
%The significance of the above formula is to introduce the rate constraint in the end-to-end model, which will be utilized as a part of the loss function in section~\ref{sec:loss}. 
In the memory decomposition phase, the reconstructed spatio-temporal feature $\bm{\hat{M}}$ can be obtained by the corresponding inverse operations.

\subsection{Clue Compression $\&$ Decompression}
We have already introduce the local information (clue) in (\ref{formula:dis}). In this paper, we utilize skeleton as the clue information for compression. It attends on memory $\bm{M}$ and helps the generation of joint representation, which is essential for frame generation in decoder. Therefore, skeleton information is also needed to compress and transmit. In this part, we design a lossless compression method to compress the skeleton. The skeleton $\bm{s_t}$ is represented by 18 body nodes and extracted by a pose estimator~\cite{cao2017realtime}. For each body node, the coordinates are used to represent the position. 

\begin{figure}[b]
    \centering
	\includegraphics[width=1.0\linewidth]{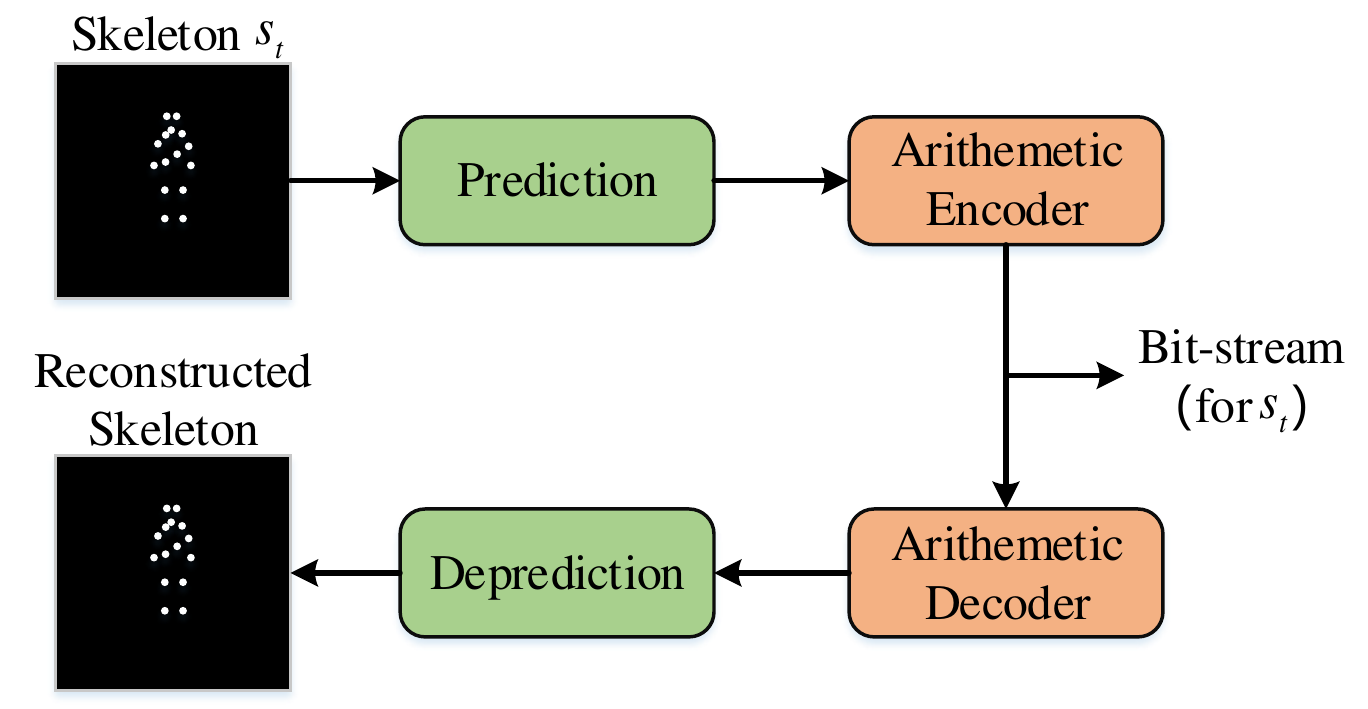}
	\caption{Our clues compression module.}
	\label{fig:cluescompress}
\end{figure}

Since there exists continuity between video frames, we first de-correlate them by predicting the coordinate $s_{t_i}$ of the $i-th$ node at the current skeleton $\bm{s_t}$ with the node in the previous skeleton $s_{{t-1}_i}$. Thereby we can calculate the residual by:
\begin{equation}
\setlength{\abovedisplayskip}{3pt}
\setlength{\belowdisplayskip}{3pt}
    res_{t_i}= s_{t_{i}}-s_{{t-1}_{i}}\label{residual}.
\end{equation}

After that, we use the adaptive arithmetic entropy coding to compress the residual information $res_{t_i}$,  which can obtain the bit-stream of the clues.  In the decompression phase, the reconstructed skeletons $\bm{\hat{s}_{1:T}}$ can be computed by the corresponding inverse operations. 

\subsection{Recall from Skeleton}
\label{sec:recall}
We formulate the combination of the spatial-temporal global information $\bm{\hat{M}}$ (memory) and local skeleton information $\bm{\hat{s_t}}$ (clue) as an attention procedure in (\ref{formula:G}). Then the output of the attention is fed into the generator to obtain reconstructed frame $\bm{x_t}$. Here we describe the detailed structure of the recalling attention mechanism $A$, the generator $g_d$ and discriminators.

\paragraph{\textbf{Recalling Attention}}

Attention mechanism has drawn considerable attention from both Natural Language Processing~\cite{bahdanau2014neural,vaswani2017attention} and Computer Vision~\cite{wang2018non,xu2015show}. Impressively, Vaswani \textit{et al.}~\cite{vaswani2017attention} introduce self-attention and encoder-decoder attention mechanisms to machine translation, achieving state-of-the-art translation performance. Similarly, Wang \textit{et al.}~\cite{wang2018non} apply self-attention to the field of computer vision as non-local neural network. The non-local neural network is able to compute the response at a position as a weighted sum of the features at all positions, instead of representing input features with a limited receptive field like convolutional neural networks. Inspired by the success of the aforementioned works, we here present \textit{Recalling Attention}, which mimics the typical recalling process existed in human behaviors. Different from non-local neural network that outputs a global representation of itself, our Recalling Attention allows the skeleton to attend over memory and generate a joint representation that combines the information from both sides. In the following, we describe our recalling attention as a function of \textit{query} and \textit{key-values} pairs.
Inspired by the success of the attention mechanism~\cite{vaswani2017attention,wang2018non}, we here present \textit{Recalling Attention}, which mimics the typical recalling process existed in human behaviors. 

%Our Recalling Attention allows clues to attend over memory and generate a joint representation that combines the information from both sides.
\begin{figure}[t]
	\centering
	\includegraphics[width=1.00\linewidth]{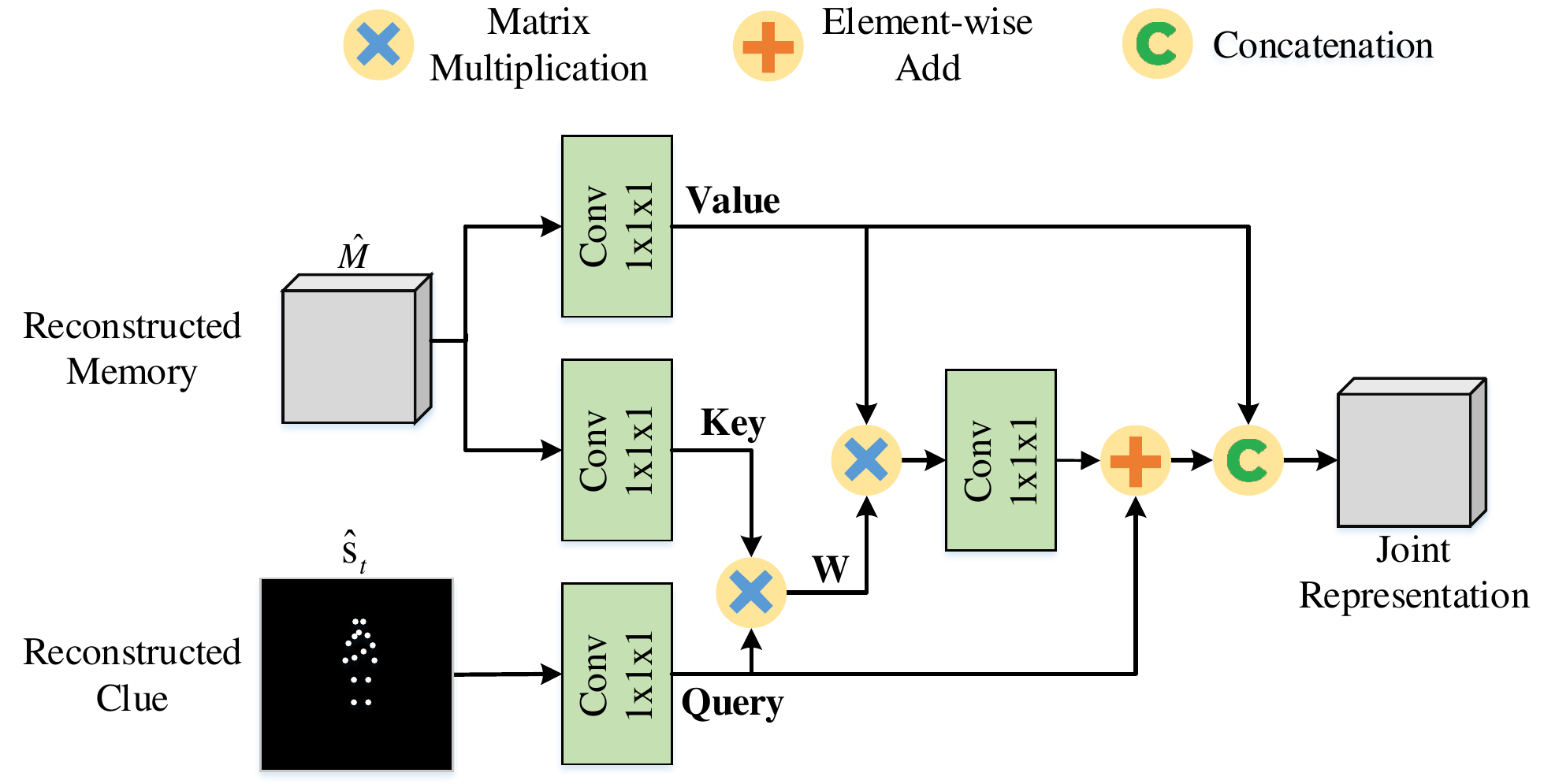}
	\caption{Our Recalling Attention module.}
	\label{fig:recallatt}
\end{figure}
Formally, we define a query matrix $Q$, a key matrix $K$ and a value matrix $V$. The Recalling Attention $R(Q, K, V)$ can be formulated as:
\begin{equation}
\begin{aligned}
    R(Q, K, V) = [W V^\mathsf{T} + Q, V],
\end{aligned}
\end{equation}
%where  "$+$" represents a residual connection, $[\cdot, \cdot]$ indicates concatenation, and $W$ is a weight matrix can be calculated as:
where $W$ is a weight matrix to be learned, "$+Q$" represents a residual connection, and $[\cdot, \cdot]$ indicates concatenation. The convolutional layer is omitted for simplicity. Note that, the weight matrix $W$ can be learned in different ways~\cite{vaswani2017attention,wang2018non}. We here adopt the simplest but effective version: 
\begin{equation}
\setlength{\abovedisplayskip}{3pt}
\setlength{\belowdisplayskip}{3pt}
    W = QK^\mathsf{T}.
\end{equation}

Adapting to our system, as Fig.~\ref{fig:recallatt} illustrated, the reconstructed spatio-temporal feature $\bm{\hat{M}}$ is regarded as key and value, and the reconstructed skeleton $\bm{\hat{s_t}}$ is regarded as the query. Intuitively, our Recalling Attention is computed as a weighted sum over memory, where the weight assigned to each part of memory is computed by the clue with the corresponding part of memory.

\paragraph{\textbf{Adversarial Generation}}
We combine the VAE and GAN by utilizing generator as the transform $g_{d}$ and replacing the distortion loss with generator loss $L_{G}$ in~\ref{sec:gan}. However, the spatial discriminator $g_{dis_S}$ only independently judges the quality of per frame, lacking the constraint for temporal continuity. To improve the temporal continuity between adjacent generated frames, we following Chan \textit{et al.}~\cite{chan2018everybody} and propose additional temporal discriminator $g_{dis_T}$ for our MTR framework. 

The difference between two discriminators can be seen in follows:
\begin{itemize}
    \item \textbf{Spatial Discriminator ($g_{dis_S}$)} takes the skeleton ($\bm{s_t}$) and the generated or real frame ($\bm{x_t}$) as input to judge whether the input frame is real or fake. The insight of Spatial Discriminator is to facilitate the generator to yield realistic images conditioned on certain input skeleton. 
    
    \item \textbf{Temporal Discriminator ($D_T$)} takes adjacent skeletons ($s_t$ and $s_{t-1}$) and corresponding generated or real frames as input to judge whether the input frames are from real video. The goal of the Temporal Discriminator is to ensure the continuity between adjacent generated frames.
    
\end{itemize} 

Based on the above design, we can modify the optimization function of discriminators with:
\begin{equation}
\begin{aligned}
L_{dis}=L_{dis_S}+L_{dis_T},
\end{aligned}
\end{equation}
where the $L_{dis_S}$ is already defined in (\ref{formula:dis}). And $L_{dis_T}$ can be written as:
\begin{equation}
\begin{aligned}
L_{dis_T}=
\mathbb{E}_{\bm{x_{1:T} \sim p_{\bm{x_{1:T}}}}}[\log \prod_{t=1}^{T} p_{d_T}(1 \mid \bm{x_{t-1:t}})]+\\
\mathbb{E}_{\bm{\hat{x}_{1:T} \sim p_{\bm{\hat{x}_{1:T}}}}}[\log \prod_{t=1}^{T} p_{d_T}(0 \mid \bm{\hat{x}_{t-1:t}})]\\
with \ p_{d_T}(1\mid \bm{x_{t-1:t}})=g_{dis_T}(\bm{x_{t-1:t}}, \bm{s_{t-1:t}}, \bm{\phi_{dis_T}}),\\
p_{d_T}(0\mid \bm{\hat{x}_{t-1:t}})=1-g_{dis_T}(\bm{\hat{x}_{t-1:t}}, \bm{s_{t-1:t}} ,\bm{\phi_{dis_T}}),
\label{formula:dis_t}
\end{aligned}
\end{equation}
where the $p_{d_T}$ is the probability obtained from temporal discriminator, predicting whether the input clip is real clip (1 for real and 0 for fake).
After that, we can evolve our generator optimization loss by:
\begin{equation}
\begin{aligned}
L_{G}=&-\mathbb{E}_{\bm{\hat{x}_{1:T} \sim p_{\bm{\hat{x}_{1:T}}}}}[\log \prod_{t=1}^{T} p_{d_{S}}(0 \mid \bm{\hat{x}_{t}})]\\
&-\mathbb{E}_{\bm{\hat{x}_{1:T} \sim p_{\bm{\hat{x}_{1:T}}}}}[\log \prod_{t=1}^{T} p_{d_T}(0 \mid \bm{\hat{x}_{t-1:t}})].
\label{formula:finalG}
\end{aligned}
\end{equation}

With the aid of spatial and temporal discriminators, we can train the generator to learn to generate video reconstructions that satisfy both single-frame authenticity and adjacent-frame continuity. In detail, we base our transformation network $g_d$ and discriminators on the objective presented in pix2pixHD~\cite{wang2018vid2vid}.

\subsection{Loss Function for End-to-end Compression Network Training}
\label{sec:loss}
In section~\ref{sec:II} and section~\ref{sec:recall}, we first optimize the compression problem through the variational inference in (\ref{formula:kl}), . Then we improve it by replacing the distortion part with generator loss in (\ref{formula:G}). After that, we utilize additional discriminator to improve the temporal continuity of the adjacent generated frames, which further improve the loss of generator $L_G$ in (\ref{formula:finalG}). Therefore, based on the above formulas, the full loss function of our model can be formulated as follows:
\begin{equation}
\ell = \lambda_{\text{rate}}R+L_G.
\label{formula:firstloss}
\end{equation}

Based on ({\ref{formula:firstloss}}), we further improve the training loss for better reconstruction quality. Following Chan \textit{et al.}~\cite{chan2018everybody}, we adopt VGG perceptual loss $\ell_{\text{VGG}}$ by adding it as a part of our distortion loss. In addition, following Ledig \textit{et al.}~\cite{ledig2017photo}, we introduce the feature matching loss $\ell_{\text{fm}}$ to improve the training process of our generative model. Therefore, the final training loss for our MTR compression network can be written as:
 
\begin{equation}
\ell = \lambda_{\text{rate}}R+L_G+\lambda_{\text{VGG}}\ell_{\text{VGG}}+\lambda_{\text{fm}}\ell_{\text{fm}},
\end{equation}
where $\lambda$ balance the importance of each part in loss funcrion. In our experiments, we heuristically set $\lambda_{\text{rate}}=1$, $\lambda_{\text{fm}}=10$ and $\lambda_{\text{VGG}}=10$ to train our MTR network.

\section{Experiments}
\begin{figure*}[t]
	\setlength{\abovecaptionskip}{0pt} 
	\setlength{\belowcaptionskip}{0pt}
	\centering
	\includegraphics[width=1.0\linewidth]{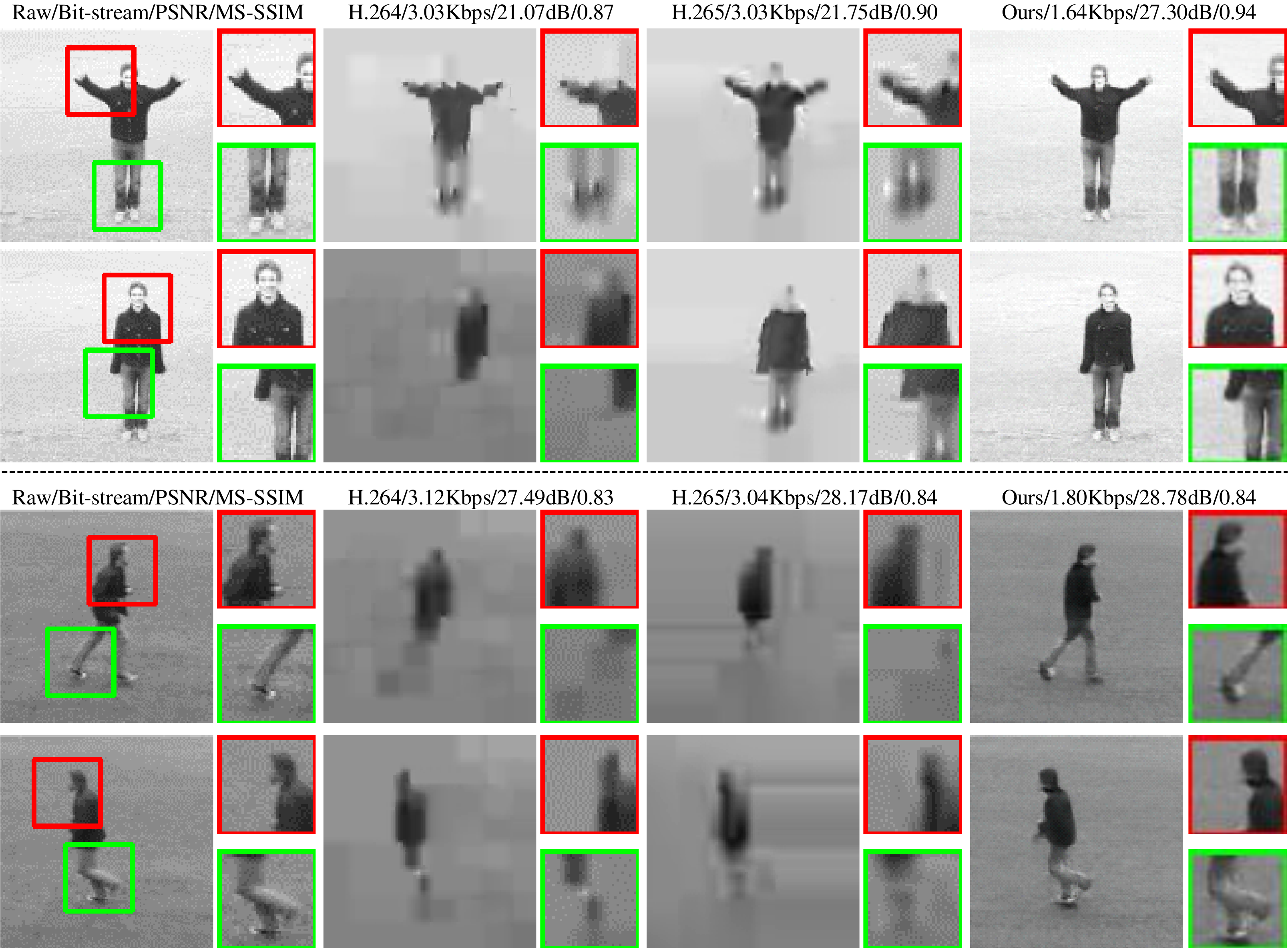}
	\caption{Comparison between our proposed method and traditional codecs on the test set of KTH dataset.}
	\label{fig:Comparsion1}
\end{figure*}

\label{sec:exps}
In this section, we first compare our MTR compression performance with traditional video compression codec. Then ablation experiments are conducted to analyze the influence of each module in our framework. Detailed settings about experiments are shown as follows:
\paragraph{\textbf{Dataset}}
%We train the proposed video compression framework using KTH dataset~\cite{laptev2004recognizing} and APE dataset~\cite{yu2013unconstrained}. We randomly divide the KTH dataset into training (130 sequences), validation (12 sequences) and test set (8 sequences) and evaluate the performance on the test set. Similarly, APE dataset is also randomly divided into training (230 sequences), validation (8 sequences) and test set (7 sequences).   
We train the proposed video compression framework using KTH dataset~\cite{laptev2004recognizing} and APE dataset~\cite{yu2013unconstrained}. KTH dataset contains six types of different human action classes: walking, jogging, running, boxing, waving and clapping. Each human action is performed several times by 25 actors, yielding 150 video sequences. Each sequence roughly contains 200--800 frames. The spatial resolution of the sequences is 160$\times$120. In our experiment, we randomly divide the KTH dataset into training (130 sequences), validation (12 sequences) and test set (8 sequences) and evaluate the performance on the test set.
The APE dataset contains 245 sequences captured from 7 actors. Video sequences of each subject are recorded in unconstrained environments, like changing person poses and moving backgrounds. Similarly, the APE dataset is randomly divided into training (230 sequences), validation (8 sequences) and test set (7 sequences).  

\begin{figure*}[t]
	\setlength{\abovecaptionskip}{0pt} 
	\setlength{\belowcaptionskip}{0pt}
	\centering
	\includegraphics[width=1.0\linewidth]{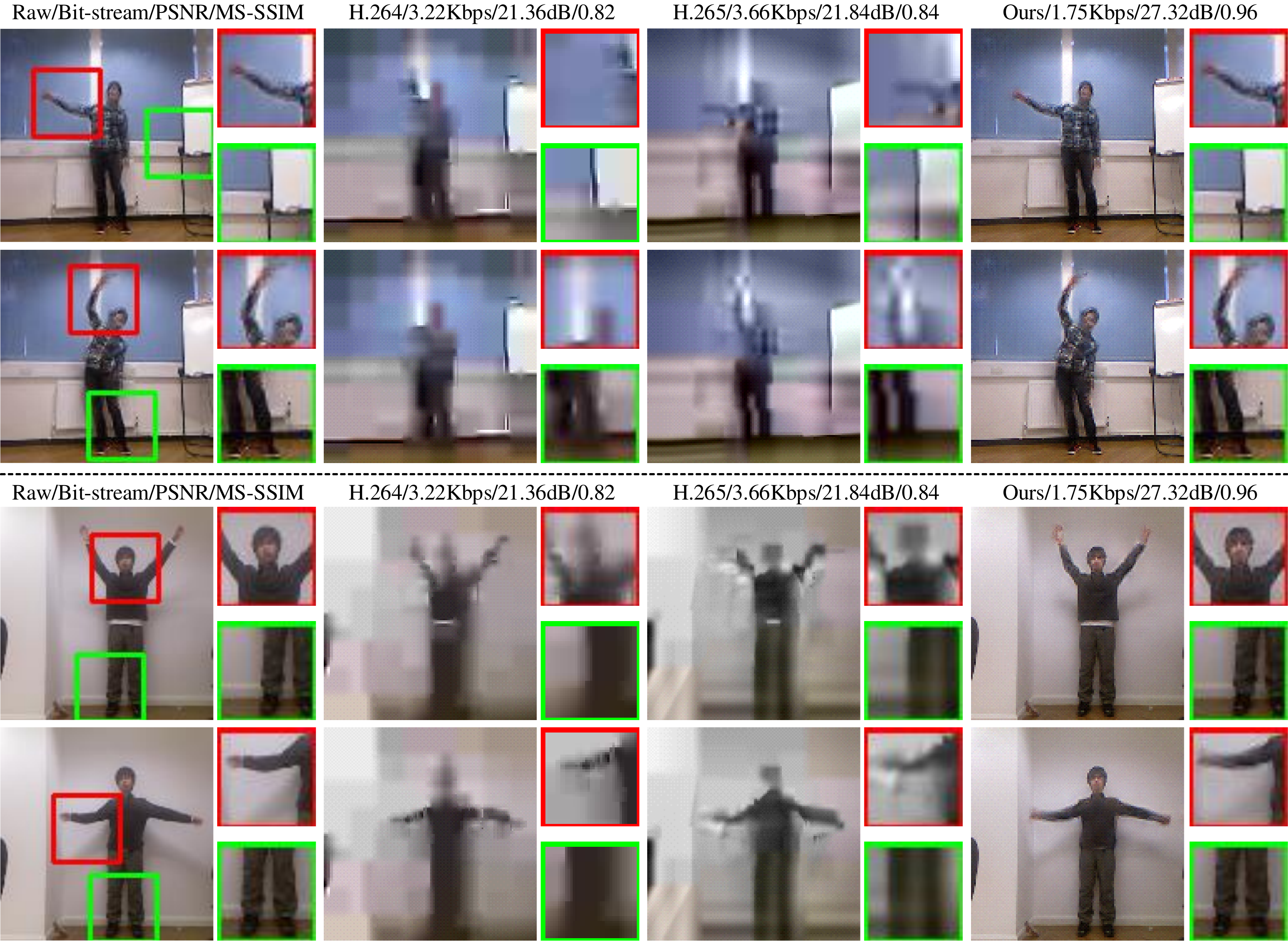}
	\caption{Comparison between our proposed method and traditional codecs on the test set of APE dataset.}
	\label{fig:Comparsion2}
\end{figure*}

\paragraph{\textbf{Implementation Details}}

We utilize the weight of the well-trained model~\cite{cao2017realtime} as the initial weight of the pose estimator. 
%Since the clue compression is a lossless compression method, we block clue compression $\&$ decompression modules in the training process. 
The output of the pose estimator is directly fed into the recalling attention module during the training phase. We utilize random crops and random horizontal/vertical flips to realize the data augmentation. The mini-batch size is 4. We use Adam optimizer~\cite{kingma2014adam} to update network parameters, in which $\beta_1$  is set as 0.5 and $\beta_2$ is 0.999. The initial learning rate is 0.0002. The whole system is implemented based on PyTorch, and it takes about one day to train the model using one NVIDIA GTX 1080Ti GPU. 

\paragraph{\textbf{Metrics}}

We adopt PSNR and MS-SSIM~\cite{wang2004image} to evaluate the performance of our scheme. PSNR is introduced as a common metric to reflect the pixel level fidelity and MS-SSIM indicates the structural similarity. The higher PSNR and MS-SSIM indicate better reconstruction quality. The bit-rate that used for transmission is denoted as Kilobits per second (Kbps) on 25 frames/second (fps).

\subsection{Comparison with Traditional Codecs}

In this subsection, we compare the compression quality of our method with the traditional video codecs, including H.264\footnote{\url{https://www.itu.int/rec/T-REC-H.264}} and H.265\footnote{\url{https://www.itu.int/rec/T-REC-H.265}}. For fairness, all codecs use the same GoP size as 10.

Fig.~\ref{fig:Comparsion1} and Fig.~\ref{fig:Comparsion2} visualizes the experimental results on the test set of KTH dataset (top two rows) and APE dataset (bottom two rows), in which the fourth column is generated by our scheme. Note that H.264 and H.265 cannot compress the sequence to a bit-rate lower than about 3 Kbps. 

Subjectively, MTR successfully generates video frames with rich details such as grassland and the colorful background. It gets rid of blocking artifacts, and preserve the reality while adapting to the specific pose. We also provide a quantitative evaluation of our framework in Table~\ref{codecresult}, from which we can see that for APE dataset, our scheme significantly outperforms the strong baselines up to $3.61$dB with only $56.70\%$ bit-rate. Moreover, our model can be generalized to KTH dataset, which has more complex scenarios (e.g., camera movement), also showing comparable results with the latest video codecs. 

Besides testing the quality of the reconstruction, we also test the encoding and decoding time on the same machine (Intel Core i7-8700 CPU / NVIDIA GTX 1080Ti GPU). For a video sequence (300 frames), our model requires 29.67s, while HEVC requires 51.33s. It should be noted that our framework is not technically optimized yet, and it can be further accelerated by model compression or the latest AI chips. 

%\begin{table}[t]
%\setlength{\abovecaptionskip}{0pt} 
%\setlength{\belowcaptionskip}{-0pt}
%\centering
%\caption{Comparison with the latest traditional codecs. Results are averaged on the test set.}
%\begin{tabular}{|l|c|c|c|c|}
%\hline
%    & Method       & Rate (Kbps) & MS-SSIM     & PSNR (dB)   \\ \hline
%KTH & JM (H.264)     & 3.96        & 0.84        & 25.78 \\ \cline{2-5}
%    & HM (H.265)     & 3.54        & 0.86        & 26.92 \\ \cline{2-5} 
%    & MTR(Ours)            & 2.10        & 0.82        & 25.68 \\ \hline
    
%APE & JM (H.264)     & 3.27        & 0.84        & 23.43 \\ \cline{2-5} 
%    & HM (H.265)     & 3.21        & 0.87        & 24.04 \\ \cline{2-5} 
%    & MTR (Ours)           & 1.82        & 0.97        & 27.65    \\ \hline
%\end{tabular}
%\label{codecresult}
%\end{table}

\begin{table}[t]
\setlength{\abovecaptionskip}{0pt} 
\setlength{\belowcaptionskip}{-0pt}
\centering
\caption{Comparison with the latest traditional codecs. Results are averaged on the test set.}
\begin{tabular}{ccccc}
\toprule[1pt]
                     & Method     & Rate (Kbps) & MS-SSIM & PSNR (dB) \\ \midrule[0.5pt]
                     {KTH} & JM (H.264) & 3.96        & 0.84    & 25.78     \\
                     & HM (H.265) & 3.54        & 0.86    & 26.92     \\
                     & MTR (Ours) & 2.10        & 0.82    & 25.68     \\ \midrule[0.5pt]
                     {APE} & JM (H.264) & 3.27        & 0.84    & 23.43     \\
                     & HM (H.265) & 3.21        & 0.87    & 24.04     \\
                     & MTR (Ours) & 1.82        & 0.97    & 27.65     \\ \bottomrule[1pt]
\end{tabular}
\label{codecresult}
\end{table}
%\begin{figure}[t]
%	\centering
%	\includegraphics[width=0.85\linewidth]{fig/exp2.png}
%	\caption{Verification on structure learning. we extract the memory from one video GoP and clues from another video respectively. Our approach is able to fuse them and synthesize new frames.}
%	\label{fig:tractable clues}
%\end{figure}

%\subsection{Synthesize with new clues information}
%In our framework, we leverage the inherent structure of video sequences by decomposing video signals into a global spatiotemporal feature and skeleton for each frame. To demonstrate the effectiveness of our decomposition, we combine memory extracted from one video GoP with new clues that come from other video sequences to synthesize video frames.  

%From Fig.~\ref{fig:tractable clues}, we can see that our scheme can synthesize the video content with memory and clues coming from different sequences, which means that our compression framework learns the human structure (pose), rather than directly copying pixels from memory information.

\subsection{Ablation Experiments}

\paragraph{\textbf{Ablation on memorizing and recalling mechanisms}}
We verify the effectiveness of memorizing and recalling mechanisms by building the framework without memorizing or without recalling respectively. Specifically, the model without memorizing is implemented as directly adopting the first frame as memory, instead of memorizing over the whole sequence. While the model without recalling is implemented as directly concatenating the reconstructed spatio-temporal feature $\bm{\hat{M}}$ and skeletons $\bm{\hat{s}_{1:T}}$, rather than performing Recalling Attention. The experimental results are demonstrated in Table~\ref{ablation_MTR}, from which we can see that the model combined with both techniques (MTR) significantly outperforms two individual baselines.

%\begin{table}[h]
%\setlength{\abovecaptionskip}{0pt} 
%\setlength{\belowcaptionskip}{-0pt}
%\centering
%\caption{Ablation on model architecture. Results are obtained on KTH dataset.}
%\begin{tabular}{|c|c|c|c|}
%\hline
%               & Rate (Kbps) & MS-SSIM & PSNR  \\ \hline
%w/o recalling  & 2.13        & 0.78    & 23.47 \\ \hline
%w/o memorizing & 3.41        & 0.78    & 23.48 \\ \hline
%MonC           & 2.16        & 0.82    & 25.53 \\ \hline
%MTR (Ours)            & 2.10        & 0.82    & 25.69 \\ \hline
%\end{tabular}
%\label{ablation_MTR}
%\end{table}

\begin{table}[t]
\setlength{\abovecaptionskip}{0pt} 
\setlength{\belowcaptionskip}{-0pt}
\centering
\caption{Ablation on model architecture. Results are obtained on KTH dataset.}
\begin{tabular}{cccc}
\toprule[1pt]
               & Rate (Kbps) & MS-SSIM & PSNR (dB) \\ \midrule[0.5pt]
w/o recalling  & 2.13        & 0.78    & 23.47     \\
w/o memorizing & 3.41        & 0.78    & 23.48     \\
MonC           & 2.16        & 0.82    & 25.53     \\ 
MTR (Ours)     & 2.10        & 0.82    & 25.69     \\ \bottomrule[1pt]
\end{tabular}
\label{ablation_MTR}
\end{table}

\paragraph{\textbf{Variants of attention mechanism}}

We conduct different Recalling Attention mechanisms in this part. Specifically, there are two possible attention directions for our Recalling Attention. The first one is ``Clues Attend on Memory" (ConM, a.k.a MTR), which is employed in our scheme. As a counterpart, the second one is ``Memory Attend on Clues" (MonC). Different with ConM, MonC utilizes $\bm{\hat{s_t}}$ as the key and value, and $\bm{\hat{M}}$ is regarded as the query. We illustrate the experimental results in Table~\ref{ablation_MTR}. The result shows that MTR achieves better performance than MonC.

% -----------------------------------------------
\section{Conclusion}
\label{sec:con}
In this paper, we propose a Memorize-Then-Recall framework for low bit-rate surveillance video compression by leveraging the inherent structure between frames. With the assistance of the variational autoencoder and generative adversarial network, the proposed framework significantly surpasses the latest coding standards. In the future, we expect to more optimization and plan to extend our framework to more complex surveillance scenarios such as traffic intersections.

%\appendices
%\section{Detailed parameters for generator and discriminators}
%We provide parameters of generator ($g_d$) and discriminators ($g_{dis_S}$ and $g_{dis_T}$) in Table.~\ref{parameters}. The .

%\begin{table*}[t]
%\centering
%\caption{Detailed parameters.}
%\begin{tabular}{cccc}
%\toprule[2pt]
%               & Rate (Kbps) & MS-SSIM & PSNR (dB) \\ \midrule[0.5pt]
%w/o recalling  & 2.13        & 0.78    & 23.47     \\
%w/o memorizing & 3.41        & 0.78    & 23.48     \\
%MonC           & 2.16        & 0.82    & 25.53     \\ 
%MTR (Ours)     & 2.10        & 0.82    & 25.69     \\ \bottomrule[2pt]
%\end{tabular}
%\label{parameters}
%\end{table*}

\section*{Acknowledgment}
This work was supported in part by NSFC under Grant U1908209, 61571413, 61632001 and the National Key Research and Development Program of China 2018AAA0101400.

\ifCLASSOPTIONcaptionsoff
  \newpage
\fi


\bibliographystyle{./bibtex/IEEEtran}
\bibliography{./bibtex/IEEEabrv,./bibtex/myref}


%\end{thebibliography}


\begin{IEEEbiography}{Michael Shell}
Biography text here.
\end{IEEEbiography}

\begin{IEEEbiographynophoto}{John Doe}
Biography text here.
\end{IEEEbiographynophoto}

%\newpage

\begin{IEEEbiographynophoto}{Jane Doe}
Biography text here.
\end{IEEEbiographynophoto}


%\vfill


\end{document}